\documentclass[prd,twocolumn,english,amsmath,amssymb,floatfix,superscriptaddress,longbibliography,nofootinbib]{revtex4-2}
\usepackage[english]{babel}
\usepackage[latin9]{inputenc}
\usepackage{graphicx}
\usepackage{multirow}
\usepackage[usenames,dvipsnames]{xcolor}
\usepackage{bm}
\usepackage{mathtools}
\usepackage{subcaption}
\usepackage{rotating}
\definecolor{oceanboatblue}{rgb}{0.0, 0.47, 0.75}
\definecolor{orange}{rgb}{1,0.5,0}
\definecolor{goodgreen}{rgb}{0.1,0.5,0}
\definecolor{goodred}{rgb}{0.7,0,0}
\usepackage[colorlinks,urlcolor=goodgreen,citecolor=blue,linkcolor=goodred]{hyperref}
\usepackage{cleveref}
\usepackage{lineno}
\usepackage{braket}
\usepackage{tikz}
\usepackage{tocvsec2}
\usepackage{scrextend}
\makeatletter

\newcommand{\comment}[1]{}

\usepackage[normalem]{ulem}

\makeatother

\def\cL{\mathcal{L}}

\def\cS{\mathcal{S}}
\def\cF{\mathcal{F}}

\def\cP{\mathcal{P}}
\def\cJ{\mathcal{J}}

\newcommand\x{\mathbf{x}}

\newcommand\+{\dagger}
\newcommand\<{\langle}
\renewcommand\>{\rangle}
\renewcommand\d{\partial}

\def\d{\partial}
\def\cL{\mathcal{L}}

\def\cN{\mathcal{N}}

\def\cA{\mathcal{A}}
\def\cB{\mathcal{B}}

\def\cS{\mathcal{S}}
\def\cN{\mathcal{N}}
\def\cF{\mathcal{F}}

\def\>{\rangle}
\def\<{\langle}

\renewcommand\d{\partial}

\newcommand\mk{\mathbf{k}}

\newcommand\mq{\mathbf{q}}

\newcommand\he{\hat{e}}
\newcommand\hw{\hat{\omega}}
\newcommand\hE{\hat{E}}
\usepackage{cleveref}

\newcommand{\be}{\begin{equation}\begin{aligned}}
		\newcommand{\ee}{\end{aligned}\end{equation}}

\newcommand{\lb}{\left}
\newcommand{\rb}{\right}

\newcommand{\eqsp}{\, ,\quad} %shrtct for space in eqns

\newcommand{\mc}{\mathcal}

\begin{document}
\title{Supergravity model of the Haldane-Rezayi fractional quantum Hall state}

\author{Dung Xuan Nguyen}
\email{dungmuop@gmail.com}
\affiliation{Center for Theoretical Physics of Complex Systems, Institute for Basic Science (IBS), Daejeon, 34126, Republic of Korea}
\author{Kartik Prabhu}
\email{kartikprabhu@ucsb.edu}
\affiliation{Department of Physics, University of California, Santa Barbara, CA 93106, USA}
\author{Ajit C. Balram}
\email{cb.ajit@gmail.com}
\affiliation{Institute of Mathematical Sciences, CIT Campus, Chennai 600113, India}
\affiliation{Homi Bhabha National Institute, Training School Complex, Anushaktinagar, Mumbai 400094, India} 
\author{Andrey Gromov}
\email{andrey@umd.edu}
\affiliation{ Department of Physics \&
  Condensed Matter Theory Center,
  University of Maryland
  College Park, Maryland 20740, USA}
\begin{abstract}
	Supersymmetry and supergravity were invented in the 1970s to solve fundamental problems in high-energy physics. Even though neither of these ideas has yet been confirmed in high-energy and cosmology experiments, they have been beneficial in constructing numerous theoretical models including superstring theory. Despite the absence of supersymmetry in particle physics, it can potentially emerge in exotic phases of strongly correlated condensed matter systems. In this paper, we propose a supergravity model that describes the low-energy physics of the Haldane-Rezayi state, a gapless quantum Hall state that occurs in a half-filled Landau level. We show that the corresponding edge modes of the Haldane-Rezayi state and the Girvin-MacDonald-Platzman algebra appear naturally in the supergravity model. Finally, we substantiate our theoretical findings with numerical exact diagonalization calculations that support the appearance of the emergent graviton and gravitino excitations in the Haldane-Rezayi state.
\end{abstract}

\maketitle
\date{\today}

\section{Introduction}
\label{sec:intro}

Supergravity, the theory that combines supersymmetry (SUSY) and general relativity, is one of the most exciting ideas in physics. It has played a central role in the development of high-energy physics in the last five decades. After the discovery of general relativity in 1915, the unification of space-time geometry with internal symmetries was the central theme that Einstein explored during the later part of his life. Supersymmetry, which interchanges bosons and fermions, was independently discovered by Gervais and Sakita in 1971 \cite{Sakita:SUSY1071}, Golfand and Likhtman in 1971 \cite{GL}, and Volkov and Akulov in 1972 \cite{Volkov:1972}. Supersymmetry provides a natural connection between the space-time and internal symmetries \cite{HAAG:1975}. Supersymmetry may explain how Higgs mass is robust to quantum corrections and solve the \textit{hierarchy problem} with a price, the appearance of superpartners of Standard Model particles. Since no superpartner has been found so far at the Large Hadron Collider, SUSY can be, at best, spontaneously broken. However, the breaking of SUSY induces massless Goldstone fermions. Supergravity, which in four dimensions was first formulated in 1973 by Volkov, and Soroka \cite{Volkov:1973} can salvage this problem. The gravitino, with spin-3/2, plays the role of a gauge field for SUSY which gets a huge mass after eating all the massless Goldstone fermions. Supergravity in higher dimensions was the first concrete realization of Einstein's dream of a unified field theory \cite{Chamseddine:GUT}. The development of supergravity was also a key contributing factor in the string theory revolution. The duality between a supergravity theory on the bulk and a conformal field theory (CFT) on the boundary, named AdS/CFT correspondence, provides a unique tool to probe quantum field theory in the strong coupling regime \cite{Maldacena:1997}. Even though supersymmetry and supergravity have not yet been confirmed in high energy physics experiments, supersymmetry can emerge in exotic phases of strongly correlated quantum matter, in particular, fractional quantum Hall (FQH) states \cite{Gunnar:MR, Palumbo:MR, Yang:MR, Gromov:SusyMR} as well as topological superconductors \cite{grover2014emergent} and cold atomic gases \cite{YuSUSY:2008,YuSUSY:2010,bradlyn2016susy}.   

The fractional quantum Hall effect (FQHE) was observed experimentally in 1982 by Tsui, Stormer and Gossard \cite{Tsui}, and the experimental findings were theoretically explained a year after by Laughlin using his eponymous wave function \cite{Laughlin:1983fy}. These discoveries lead to Nobel Prizes for Tsui, Stormer, and Laughlin. To this date, FQHE remains one of the most fertile playgrounds to probe the physics of strong correlations in condensed matter systems. Indeed, the discovery of FQHE opened a new field of research, the topological order in strongly correlated systems, of which FQHE is the simplest example, and the only one confirmed experimentally in condensed matter \cite{Wen-topo}. FQHE is realized when a two-dimensional electron gas is placed in a large perpendicular magnetic field. The magnetic field turns the Fermi-liquid of electrons into highly degenerate bands known as Landau levels (LLs) and FQHE arises when a LL is partially filled. The magnetic field completely quenches the kinetic energy of the electrons and the physics is entirely dictated by the Coulomb repulsion between the electrons. This setting realizes a wide variety of topological phases depending on the filling fraction and the effective interaction between electrons, which in turn is determined by the LL that is partially occupied. Consequently, many theoretical models have been proposed to explain the different phases realized in FQH systems \cite{Anyon1, Anyon2, ISO:1992, Jain1}. Certain FQH states can host non-abelian anyonic excitations that can form building blocks of a fault-tolerant topological quantum computer \cite{Nayak:TopoQ}. The FQH states also serve as a potential avenue to look for emergent graviton(s) in a condensed matter system \cite{Haldane2011, Golkar:2016a, Golkar2016, Liu20, Nguyen:2021a, Balram21d, Wang21a}. 

In this paper, we propose a $\cN=(1,1)$ supergravity model that describes the low-energy physics of the Haldane-Rezayi (HR) state ~\cite{HR:1988, HR:1988Erratum} which occurs at the filling fraction $\nu=1/2$ and is believed to be a compressible state~\cite{GurarieHR:1997, Read00, Seidel11, Crepel19}. Our model predicts the existence of a spin-3/2 excitation, the gravitino, in addition to the spin-2 graviton, which is the long-wavelength limit of the magnetoroton excitation that is expected to appear in every FQH state \cite{GMP:1986}. We connect the supergravity model with the bimetric theory of FQH states \cite{Gromov2017} and obtain the Girvin-MacDonald-Platzman (GMP) algebra in the long-wavelength limit \cite{GMP:1986}. We also show that our bulk Chern-Simons (CS) theory leads exactly to the conformal field theory (CFT) on the boundary that is known to describe the HR state \cite{Milovanovic:1996nj, GurarieHR:1997}. In addition, we provide numerical evidence in support of the existence of the emergent graviton and gravitino excitations. Our paper further reinstates that FQHE provides an ideal setting to study exotic theoretical models.    	 
%%=======================================================================
\section{ Supergravity model of Haldane-Rezayi state}
\label{sec:Action}
In this section, we will present the superalgebra and propose the supergravity action that describes the low-energy physics of the Haldane-Rezayi state. In our construction, the graviton corresponds to the long-wavelength limit of the magnetoroton excitation, and the gravitino represents the long-wavelength limit of the neutral fermion excitation. 

%%------------------------------------------------------------
\subsection{Non-relativistic $\cN=(1,1)$ Superalgebra}
\label{sec:susy-alg}

We will consider a $\cN=(1,1)$ supergravity model with the SUSY algebra given by:
\begin{align}
\label{eq:algebra0}
\lb[\cJ, \mc P_a\rb] = \epsilon_a{}^b \cP_b, \,\,\, \lb[Q_\alpha , Q_\beta \rb] = \gamma^a_{\alpha\beta} \cP_a, \,\,\,  \lb[\cJ , Q_\alpha \rb] = \tfrac{1}{2} \hat\gamma^\beta{}_\alpha Q_\beta.
\end{align}
Here \(\cJ\) is the generator of spatial rotations, \(\cP\) is the generatorof spatial translations and \(Q_\alpha\) is the generator of supersymmetries. In \cref{eq:algebra0}, $a,b=1,2$ are the spatial indices, and $\alpha,\beta=1,2$ denote the spinor indices. Our conventions on the Dirac matrices $\gamma$ are given in \cref{sec:dirac-alg}. The super bracket used in \cref{eq:algebra0} is defined as 
\begin{equation}
	[A,B]=AB-(-1)^{F_A F_B}BA,
\end{equation}
where $F_A$ is the fermion number of operator $A$. The relation of the algebra stated in \cref{eq:algebra0} to the usual $\cN=(1,1)$ superalgebra is described in \cref{sec:dirac-alg}.

Let $\mc T$ be the generator of background electromagnetic \(U(1)\) transformations which commutes with all the other generators. The bilinear invariants of the central extended superalgebra are given by
\be
\label{eq:invariants}
\langle\mc J \mc J \rangle = \mu_1 \eqsp \langle\mc T \mc T \rangle = \mu_A \eqsp \langle\mc J \mc T \rangle= \mu_B, \\ \langle\mc P_a \mc P_b \rangle= \mu_2\delta_{ab} \eqsp \langle Q_{\alpha} Q_{\beta} \rangle =i \mu_3 \epsilon_{\alpha\beta}.
\ee
and $\epsilon_{\alpha\beta}=i\sigma_y$. The coefficients \(\mu_1, \mu_2, \mu_3, \mu_A, \mu_B\) are constants and hence also invariant under the action of the superalgebra. These invariants are used to define the supertrace on the basis of the superalgebra. The computation of the supertrace in our action is given in \cref{sec:derivation}.

We then introduce the background $U(1)$ electromagnetic connection 
\begin{equation}
	\cA_\mu=A_\mu \mc T,
\end{equation} 
with $A_\mu$ being the background electromagnetic field, and the supergravity connection 
\begin{equation}
\label{eq:Bmu}
	\cB_\mu=\hw_\mu \cJ +\he^a_\mu \mc P_a +\psi^\alpha_\mu Q_\alpha.
\end{equation}
The field components of the supergravity connection $\cB$ are dynamical fields that describe the low-energy physical degrees of freedom of the HR state. The SUSY transformations of the connections are given by
\begin{align}
	\label{eq:SUSYA}
	\delta_\xi \cA&=0\\
	\label{eq:SUSYB}
	\delta_\xi \cB&=d\xi^\alpha Q_\alpha+ \lb[\cB,\xi^\alpha Q_\alpha \rb]
\end{align}
with the infinitesimal SUSY transformation $\xi^\alpha Q_\alpha$.
From the explicit SUSY transformation given in \cref{eq:SUSYB}, we see that 
\be
\label{eq:SUSYtrans}
\delta_\xi \hw_\mu=0 \eqsp \delta_\xi \he^a_\mu = \gamma^a_{\alpha\beta} \psi^\alpha_\mu \xi^\beta \eqsp \\\delta_\xi \psi^\alpha_\mu = \partial_\mu \xi^\alpha + \tfrac{1}{2} \hw_\mu \hat\gamma^\alpha{}_\beta \xi^\beta.
\ee
As we will see, the invariance of the spin connection under supersymmetry transformations implies the SUSY invariance of the charge density operator.
We define the super field strength
$
	\cF=d\cB+\lb[\cB,\cB\rb]
$
with its explicit form in terms of field components being
\be
\cF = (d\hw) \cJ +T^a \mc P_a + T^\alpha Q_\alpha,
\ee
and the definition of torsion is
\be 
\label{eq:torsion}
T^a &=d \he^a+	(\hw \epsilon_b{}^a) \wedge \he^b + \gamma^a_{\alpha\beta}\psi^\alpha \wedge \psi^\beta, \\
T^\alpha &= d\psi^\alpha + \tfrac{1}{2} \hat\omega \hat\gamma^\alpha{}_\beta \wedge \psi^\beta.
\ee
% The SUSY transformation of the torsion is \KP{do we need this?}
% \be
% \delta_\xi T^a = \gamma^a_{\alpha\beta} T^\alpha \xi^\beta \eqsp \delta_\xi T^\alpha = \tfrac{1}{2} (d\hw) \hat\gamma^\alpha{}_\beta \xi^\beta
% \ee
% %
The Bianchi identity is \(0 = d\cF + \lb[ \cB, \cF \rb]\) which in components becomes
\be
0 & = dR = dd\hw \\
0 & = dT^a + (\hw \epsilon_b{}^a) \wedge T^b - (R \epsilon_b{}^a) \wedge e^b + \gamma^a_{\alpha\beta} \psi^\alpha \wedge T^\beta \\
0 & = dT^\alpha + \tfrac{1}{2} (\hw \hat\gamma^\alpha{}_\beta) \wedge T^\beta - \tfrac{1}{2} (R \hat\gamma^\alpha{}_\beta) \wedge \psi^\beta
\ee

%We explicitly have the SUSY variation of CS term \(\psi_\alpha \wedge T^\alpha\):
%\be
%\delta_\xi \psi_\alpha \wedge T^\alpha + \psi_\alpha \wedge \delta_\xi T^\alpha = d(\xi_\alpha T^\alpha) + R \wedge (\hat\gamma_{(\alpha\beta)}\xi^\alpha\psi^\beta)
%\ee
%and the SUSY variation of CS term \(e_a \wedge T^a\):
%\be
%\delta_\xi e_a \wedge T^a + e_a \wedge \delta_\xi T^a 
%&= \gamma^a{}_{\alpha\beta} \lb[\psi^\alpha \wedge T_a + e_a \wedge T^\alpha \rb] \xi^\beta 
%\ee
%

%%%%%%%%%%%%%%%%%%%%%%%%%%%%%%%%%%%%%%%%%%%%%%%%%%%%%%%%%%%%%%%%%%%%%%%%%%%%%%
\subsection{The supergravity action}

%\KP{explain the drift velocity frame somewhere here}

Employing the extended superalgebra in the previous section, we will construct the supergravity model that captures the low-energy dynamics of the HR state. We consider the Chern-Simons (CS) action that couples the background electromagnetic field with the supergravity sector
\be\label{eq:Ls0}
\cL_{CS} =\text{sTr} \bigg(&\frac{\nu}{\mu_A 4\pi}\cA \wedge d\cA+\frac{\alpha}{\mu_B} \cA \wedge d \cB + \cB \wedge d\cB \\
& + \frac{2}{3}\cB \wedge \cB \wedge \cB \bigg)
\ee
where $ \text{sTr}$ is the super trace with the property
\begin{equation}
	\label{eq:str}
	\text{sTr}(A B)=(-1)^{F_A F_B}\text{sTr}(B A)
\end{equation}
We see that the action of \cref{eq:Ls0} is SUSY-invariant when the super field strength satisfies $\cF=0$ i.e., on-shell.\footnote{To satisfy SUSY off-shell, one would need to add auxiliary fields to the action. We leave this approach for future work. } 
Using the algebra of \cref{eq:algebra} and the bilinear invariants given in \cref{eq:invariants}, we can rewrite the action in terms of field components. The explicit form of the supergravity CS action is 
\be
\label{eq:Ls1}
	\cL_{CS} & =\frac{\nu}{4\pi} A \wedge dA+\alpha A \wedge d\hw+ \mu_1 \hw \wedge d \hw \\
&\quad + \frac{\mu_2}{2} \delta_{ab}\he^a \wedge T^b + i\mu_3 \epsilon_{\alpha \beta}\psi^\alpha \wedge T^\beta .
\ee
The equation of motion of $\he^a$ gives us the zero torsion constraint
\begin{equation}
	\label{eq:Torsion0}
	T^a=d \he^a+	(\hw \epsilon_b{}^a) \wedge \he^b + \gamma^a_{\alpha\beta}\psi^\alpha \wedge \psi^\beta=0\,.
\end{equation}
In the original bimetric theory for FQH states \cite{Gromov2017}, there is no $\he^a \wedge T^b$ term, and $\hw$ was defined in term of $\he$ through a torsion-free condition. Therefore we drop the $\he^a \wedge T^b$ term\footnote{By imposing the bilinear invariant $\langle \cP \bar{\cP}\rangle=0$.} but keep the torsion-free condition of \cref{eq:Torsion0} to replace $\hw$ in terms of $\he$ and $\psi$. 

We define the (pseudo-)inverse vielbein $\hE_a^\mu$ with the following properties
\begin{equation}
	\hE^\mu_a \he^a_\nu = \delta_\nu^\mu, \quad \hE^\mu_a \he^b_\mu = \delta^b_a. 
\end{equation}
Rectangular vielbeins appeared in effective theories of FQH in \cite{Bradlyn-Read, Gromov-Abanov, Gromov-Bradlyn-Scott}.
The torsion-free condition \eqref{eq:Torsion0} can be solved to write the emergent spin connection in terms of the emergent vielbein. The spatial component of emergent spin connection is given by 
\begin{equation}
	\label{eq:wi}
	\hw_i=- \epsilon_{ij}\epsilon^{kl}\epsilon_b{}^a \left( \hE^j_a \d_k \he^b_l + \hE^j_a \gamma^b_{\alpha\beta} \psi^\alpha_k \psi^\beta_l\right).
\end{equation}
As we show in \cref{sec:RS} that the gravitino contributions to the above solution vanish when the Rarita-Schwinger gauge is imposed. Using the gauge $\psi^\alpha_t=0,$ on the non-dynamical spinor field, we obtain the explicit expression of the time component of the emergent spin connection
\be
	\label{eq:w0}
	\hw_t = \frac{1}{2} \epsilon_b{}^a \hE^i_a \lb[ \d_t \he^b_i - \lb( \d_i \he^b_t + \epsilon_c{}^b \hw_i \he^c_t \rb) \rb].
\ee

We then arrive at the CS supergravity action 
\be
	\label{eq:S0}
	\cL_{CS}=&\frac{\nu}{4\pi} A \wedge dA+\alpha A \wedge d\hw+ \mu_1 \hw \wedge d \hw \\&+ i\mu_3 \epsilon_{\alpha \beta}\psi^\alpha \wedge T^\beta .
\ee
with the explicit form of the spin connection given in \cref{eq:w0,eq:wi}. In the Lagrangian of \cref{eq:S0}, $\nu=1/2$ is the filling fraction of the HR state and the first term of \eqref{eq:S0} gives the Hall conductance $\sigma_H=1/2$ in units of $e^2/h$. We will see in the next section that the corresponding boundary theory of the CS supergravity action of \cref{eq:S0} is the CFT of edge modes of the HR state. 

Some comments are in order. In comparison with the bimetric theory of FQH states \cite{Gromov2017}, the bosonic sector is similar to an emergent spin connection that describes the dynamical spin-2 excitation. This excitation is the long-wavelength limit of the magnetoroton excitation, the universal collective mode in FQH systems proposed by Girvin, MacDonald, and Platzman \cite{GMP:1986} as the lowest LL (LLL) projected charge density wave. The magnetoroton was shown to have spin-2 in the long-wavelength limit and reinterpreted as a dynamical emergent metric \cite{Golkar2016, Haldane2011, Gromov2017}, the FQH graviton. The last term of \cref{eq:S0} belongs to the fermionic sector, which represents the dynamical spin-3/2 mode, the gravitino excitation of the HR state. As we have seen, the bosonic and fermionic sectors transform to each other under the SUSY transformation of \cref{eq:SUSYtrans}.

%%%%%%%%%%%%%%%%%%%%%%%%%%%%%%%%%%%%%%%%%%%%
\subsection{Rarita-Schwinger gauge}
\label{sec:RS}
The emergent dynamical metric is defined through the emergent vielbein as follows \cite{Gromov2017}
\begin{equation}
\label{eq:metric}
	\hat{g}_{ij}= \delta_{ab} \he^a_i \he^b_j \delta_{ij}.
\end{equation}
The graviton dynamics is described by the emergent metric with the unimodular constraint\footnote{The constraint $\det{\hat{g}}=1$ removes the spin-$0$ dilaton degree of freedom, leaving only the spin-$2$ degrees of freedom.}
\begin{equation}
	\det{(\hat{g}_{ij})}=1,
\end{equation}
which in turn implies 
\begin{equation}
	\det{(\he^a_i)}=1. 
\end{equation}	
Using the SUSY transformation of vielbein given in \cref{eq:SUSYtrans}, we see that the constraint $\det{(\he^a_i)}=1$ is SUSY invariant up to leading order of perturbation\footnote{We consider $\hat{e}^a_i=\delta^a_i+\delta \hat{e}^a_i $, where $\delta \hat{e}^a_i$ is treated as a perturbation on the flat background space-time. } if the spinor field $\psi^\alpha_i$ satisfies the Rarita-Schwinger (RS) constraint\footnote{This constraint comes from $\delta_\xi \he^a_a=0 $ which is required to preserve the constraint $\det{(\he^a_i)}=1$ under SUSY transformations.} 
\begin{equation}
	\gamma^\mu_{\alpha \beta}\psi^\alpha_\mu = 0\,, 
\end{equation}
which is nothing but the constraint for spin-$\frac{3}{2}$ degree of freedom to be described by $\psi^\alpha_i$ \cite{Freedman:SUGRA}.
Using the explicit form of the gamma matrices given in \cref{eq:gamma} and the Majorana condition (see \cref{sec:dirac-alg}) on the spinor field, the RS condition implies $\psi_x=i \psi_y$. The RS condition and the explicit form of the gamma matrices imply that the contribution from spinor field $\psi^\alpha_i$ in the emergent spin connection of \cref{eq:wi} are canceled by the anti-commutativity of Grassmann variables. We can solve for \(\hw_i\) in terms of the vielbein to get 
\begin{equation}
	\label{eq:wi1}
	\hw_i=- \epsilon_{ij}\epsilon^{kl}\epsilon_b{}^a \hE^j_a \d_k \he^b_l\,. 
\end{equation}
One can see from the above argument that the RS constraint on the fermion field is nothing but the SUSY complement of the unimodular constraint on the vielbein. The RS constraint has been used to construct the spin-3/2 field in other contexts \cite{RaritaSchwinger, Luttinger:1956, Nguyen:Raman}, in particular, it was used to construct the gravitino field in various supergravity models \cite{VANNIEUWENHUIZEN1981, Freedman:SUGRA, DESER1983, DESER1984}. One consequence of the RS constraint and \cref{eq:wi1} is that the emergent gravitino $\psi^\alpha_\mu$ is a neutral fermion that does not couple directly to the background electromagnetic field $A_\mu$. However, it can couple to $A_\mu$ indirectly through the emergent graviton. 

%-------------------------------------------------------------------------------------

%-----------------------------------------------------------------------------------
\subsection{Charge density operator and the GMP algebra}  
\label{sec:GMP}
The GMP algebra \cite{GMP:1985, GMP:1986} determines the commutation relation of the LLL-projected charge density operator at different wavelengths
\be
\label{eq:GMP0}
	[\rho(\mk),\rho(\mq)]=2i e^{\frac{1}{2}(\mk \cdot \mq)}\sin\left(\frac{\mk \times \mq}{2}\ell^2_B\right)\rho(\mq),
\ee
where the magnetic length $\ell_B=1/\sqrt{B}$. The GMP algebra was shown to be equivalent to the $W_\infty$ algebra discovered in the string theory context \cite{CAPPELLI:1993,ISO:1992}. In this section, we will show that our model satisfies the GMP algebra in the long-wavelength limit.

From the action of \cref{eq:S0}, we obtain the charge density operator by variation of the action with respect to the scalar potential 
\begin{eqnarray}
	\label{eq:rho}
	\rho= \frac{\delta \cS}{ \delta A_0}=\frac{\nu B}{2\pi} + \frac{\alpha}{2} \epsilon^{ij} \hat{R}=\bar{\rho}+ \delta \rho,
\end{eqnarray}
where $\bar{\rho}=\frac{\nu B}{2\pi}$ is the average charge density of the FQH state. The emergent Ricci curvature is given by 
\begin{eqnarray}
	\hat{R}=\frac{2}{\sqrt{\hat{g}}}\epsilon^{ij} \d_i \hw_j,
\end{eqnarray}
where the emergent metric satisfies the unimodular constraint $\det(\hat{g}_{ij})=1$. Eq.~\eqref{eq:rho} relates the charge density to the emergent curvature, it connects the magnetoroton excitation, the charge density wave, to the emergent graviton. Furthermore, we see that the charge density is invariant under the SUSY transformation of \cref{eq:SUSYtrans}. 

The coupling of the background magnetic field with the time component of the emergent spin connection in the action is given by the term \(\alpha B \hat \omega_0\). Using the solution given in \cref{eq:w0} we see that only the time derivative of \(\he^a_i\) arises in this term and \(\he^a_t\) only has spatial derivatives.\footnote{Since no time derivatives of \(\he^a_t\) contribute to the action, this component of the emergent vielbein can be thought of as a Lagrange multiplier.} This implies the canonical commutation relations
\begin{equation}
	\label{eq:eE}
	[\hE^i_a(\x), \he^b_j(\x')]=-\frac{i}{\alpha} \epsilon_a{}^b \delta^i_j\delta(\x-\x').
\end{equation}
With the commutation relation of \cref{eq:eE} and the definition of charge density given in \cref{eq:rho} in terms of the emergent spin connection, one can obtain the long-wavelength limit of the GMP algebra \cite{Gromov2017, Nguyen2018}
\begin{eqnarray}
\label{eq:GMP}
	[\delta \rho(\mathbf{k}), \delta \rho(\mathbf{q})]=i \ell_B^2 (\mathbf{k} \times \mathbf{q}) \delta \rho (\mathbf{k+q})
\end{eqnarray} 
The algebra of \cref{eq:GMP} is the classical version of the $W_\infty$ algebra named $w_\infty$. It is also the algebra of area-preserving diffeomorphisms that should be fulfilled by a proper effective theory of FQHE \cite{Du:2022}.

%%%%%%%%%%%%%%%%%%%%%%%%%%%%%%%%%%%%%%%%%%%%%%%%%%%%%%%%%%%%%%%%%%%%%%%%%%%%%%%%%%%%%%%%

\section{The boundary theory}
\label{Boundary theory}
In this section, we derive the boundary theory associated with the dynamical emergent spin connection \(\hw\) and the gravitino \(\psi^\alpha\) given by the action stated in \cref{eq:S0} in the absence of perturbation of the background electromagnetic field, i.e. magnetic field $B=\bar{B}$ is a constant, and no applied electric field $\mathbf{E}=0$.
\subsection{Bosonic sector equations of motion}

From the action given in \cref{eq:S0}, we obtain the equations of motion for the dynamical fields. Varying the action with respect to the gravitino field gives us that the fermionic torsion vanishes, i.e.,
\be
  T^\alpha = 0 \,.
\ee
In the bosonic sector, if we vary the action with respect to the emergent spin connection we get
\be
  \delta_{\hw} S = \int d^3x ~ U^\mu \delta \hw_\mu \eqsp U^\mu = - \tfrac{1}{2} \varepsilon^{\mu\nu\lambda} \lb( \tfrac{1}{2}\alpha F_{\nu\lambda} + \mu_1 \hat{R}_{\nu\lambda} \rb).
\ee
Now we need to convert the variation of the spin connection to a variation of the vielbein. This can be done using \cref{eq:w0,eq:wi1} or by directly using the results of Ref.~\cite{Geracie:2016dpu}. The final result is
\be
  \delta_{\he} S = 2 \int d^3x~ \lb( \varepsilon^{\lambda (\mu} \nabla_\lambda U^{\nu)} \hat{E}_{a\nu}\rb)\delta \he^a_\mu.
\ee
Thus the equations of motion are
\be
\label{eq:eom1}
  \nabla_i(\alpha B + 2 \mu_1 R) = 0 \eqsp \nabla_{(i} \hat R_{j)t} - 2 \delta_{ij} \nabla^k \hat{R}_{kt} = 0,
\ee
where we have used the absence of electric field $\mathbf{E}=0$. Taking a trace, the second one simplifies to
\be
  \nabla^i \hat R_{it} = 0 \eqsp \nabla_{(i} \hat R_{j)t} = 0.
\ee
Thus the only degree of freedom in the curvature is
\be
\label{eq:eom2}
  \nabla_{[i} \hat R_{j]t} = - \tfrac{1}{2} \varepsilon_{ij} \d_t \hat R,
\ee
which follows from the Bianchi identity. From \cref{eq:eom1}, the equation of motion of vielbein in the absence of perturbation of the external electromagnetic field requires a constant emergent curvature $\hat{R}$ in space at any given time. We choose this constant to equal zero since from \cref{eq:rho}, the total electric charge of a FQHE state in a homogeneous magnetic field should be $\mathcal{Q}=\text{Area}\times\rho=\text{Area}\times\nu\frac{\bar{B}}{2\pi}$. Therefore, the equation of motion of the bosonic sector requires spin connection to be a pure spatial rotation 
\begin{equation}
  \hw_i=\partial_i \varphi.
\end{equation}

\subsection{Boundary action}
We consider the manifold \(M\) where the spatial region is the upper half-plane \(y \geq 0\) and the boundary \(\d M\) is at \(y=0\) coordinatized by \(t\) and \(x\).\footnote{A similar analysis can be done with a circular spatial boundary instead of a line.}

Take the action
\be
S = \int_M \alpha \bar{A} \wedge d\omega +\mu_1 \int_M \hw \wedge d \hw + i\mu_3 \int_M~ \epsilon_{\alpha\beta} \psi^\alpha \wedge T^\beta.
\ee
The variation of the action is then
\be
\delta S &= \int_M\alpha \bar{B} \wedge \delta \omega+2 \mu_1 \int_M d\hw \wedge \delta \hw + 2i\mu_3 \int_M \epsilon_{\alpha\beta} \delta\psi^\alpha \wedge T^\beta 
  \\&\quad - \mu_1 \int_{\d M} \hw \wedge \delta \hw - i\mu_3 \int_{\d M}~ \epsilon_{\alpha\beta} \psi^\alpha \wedge \delta \psi^\beta .
\ee
The bulk terms give the equations of motion as discussed before. For the boundary terms to vanish, we take the boundary conditions
\be\label{eq:bcs}
(\hw_t - v_b \hw_x )\big\vert_{\d M} = 0 \\
(\psi^\alpha_t - v_f \psi^\alpha_x)\big\vert_{\d M} = 0,
\ee
where \(v_b\) and \(v_f\) are the velocities of the bosonic and fermionic boundary modes respectively.

To obtain the boundary action, we consider solutions that satisfy the initial data constraints from the equations of motion. For the bosonic sector, we have \(\hw_i = \d_i \varphi\), i.e., the spin connection is a pure spatial rotation gauge mode.\footnote{In terms of the emergent vielbein this mode corresponds to a local rotation of the flat vielbein: \(\he^1_x = \hE_1^x = \he^2_y = \hE_2^y = \cos\varphi\) and \(-\he^1_y = -\hE_1^y = \he^2_x = \hE_2^x = \sin\varphi\).} For the gravitino, the constraint is that the spatial part of the torsion vanishes, \(T^\alpha{}_{ij} = 0\) whose solution is given by
\be\label{eq:psi-soln}
\psi^\alpha_i = D_i \lb( \lb(e^{-\tfrac{1}{2} \varphi \hat\gamma}\rb)^\alpha{}_\beta  \chi^\beta\rb) = \lb(e^{-\tfrac{1}{2} \varphi \hat\gamma}\rb)^\alpha{}_\beta \d_i \chi^\beta,
\ee
where \(D\) is the covariant derivative of the spinor field with respect to the emergent spin connection and the right-hand side is interpreted as a matrix exponential. It is convenient to write this in the Weyl basis where \(\hat\gamma\) is diagonalized (see \cref{sec:dirac-alg}), which gives
\be\label{eq:psi-soln-weyl}
\psi^\alpha_i = \begin{pmatrix}
e^{-\tfrac{i}{2} \varphi} \d_i \chi \\
e^{+\tfrac{i}{2} \varphi} \d_i \bar\chi
\end{pmatrix}.
\ee

Using the constraints, we rewrite the CS action as
\be
S & = \mu_1 \int_M d^3x \epsilon^{ij} \lb[- \hw_i \d_t \hw_j + \hw_t \d_i \hw_j + \hw_i \d_j \hw_t \rb] \\
&\quad + i\mu_3 \int_M d^3x \epsilon^{ij} \epsilon_{\alpha\beta} \lb[ - \psi^\alpha_i D_t \psi^\beta_j + \psi^\alpha_i D_j \psi^\beta_t \rb].
\ee
Since the emergent spin connection is flat, we can write the above as a boundary term after integration-by-parts and use the constraints and the boundary conditions [\cref{eq:bcs}], to get
\be
S &= \mu_1 \int_{\d M} dtdx \lb[ \d_x \varphi \d_t\varphi - v_b (\d_x \varphi)^2 \rb] \\
&\quad + i\mu_3 \int_{\d M} dt dx \lb[ \d_x \chi \d_- \bar\chi - \d_x \bar\chi \d_-\chi \rb],
\ee
where \(\d_- = \d_t - v_f \d_x\). The chiral bosonic action is the Floreanini-Jackiw action \cite{Floreanini:1987}. Since for fermionic fields \((AB)^\dagger = \bar B \bar A\) the fermionic action is the same as the HR boundary action given in Milovanovi\'c and Read \cite{Milovanovic:1996nj} and in Gurarie \emph{et al}.~\cite{GurarieHR:1997}. Notice that in Ref. \cite{Ma:EdgeMR}, the boundary theory of the Moore-Read state was constructed using $\cN=(1,0)$ supersymmetry, the boundary theory of \cite{Ma:EdgeMR} has a chiral boson and a copropagating Majorana fermion. In contrast, we consider $\cN=(1,1)$ supersymmetry in our model, as a consequence, the boundary theory includes a chiral boson and a complex fermion (even though our bulk gravitino is Majorana from the point of view of the bulk spatial geometry). %\KP{review this change in wording, and delete note if correct} 

The physical model of the edge of quantum Hall liquid was proposed by Wen \cite{Wen:Edge}, in which the edge modes are chiral Luttinger liquids. Assuming bulk-edge correspondence holds, the boundary theory is the effective theory that relates to the topological properties of the bulk, and the electron wave function in bulk can be constructed from the correlations of the effective edge theory. However, in realistic quantum Hall systems, one needs to add perturbations, including the interactions between edge modes, to the boundary theory. The perturbations depend on the physical details of the confining potential and the interactions between electrons. The modifications of the edge theory lead to quantum phase transitions that describe the quantum Hall edge reconstructions \cite{Yang:Edge2003} that were suggested in experiments \cite{Grayson:Edge1998, Chang:Edge2001, Hilke:Edge2001}.

%%==================================================================
\section{Numerical confirmation of the emergent graviton and gravitino}
\label{sec:Num}

The Haldane-Rezayi (HR) state~\cite{HR:1988, HR:1988Erratum} is described by the wave function
\begin{eqnarray}
        \label{eq: Haldane_Rezayi_wf}
	\Psi^{\rm HR}_{1/2} &=& {\rm Det} \left( \frac{1}{\left(z^{\uparrow}_{i}-z^{\downarrow}_{j}\right)^{2}}\right) \prod_{i,j} \left(z^{\uparrow}_{i} - z^{\downarrow}_{j} \right)^{2} \\
	& \times & \prod_{i<j} \left(z^{\uparrow}_{i} - z^{\uparrow}_{j}\right)^{2}\prod_{i<j} \left(z^{\downarrow}_{i} - z^{\downarrow}_{j}\right)^{2}
	e^{- \sum_{i} \left( \frac{|z^{\uparrow}_{i}|^{2}+|z^{\downarrow}_{i}|^{2}}{4\ell_{B}^{2}} \right)} \nonumber,
\end{eqnarray}
where ${\rm Det}$ stands for determinant, $z^{\uparrow}_{i}$ denotes the two-dimensional coordinate of the $i^{\rm th}$ electron with spin up $|{\uparrow}\rangle$ parametrized as a complex number with $z{=}x{-}iy$, $z^{\downarrow}_{j}$ denotes the coordinate of the $j^{\rm th}$ electron with spin down $|{\downarrow}\rangle$. The HR wave function of Eq.~\eqref{eq: Haldane_Rezayi_wf} can be constructed for an \emph{even} number of particles and describes an $S{=}0$ spin-singlet state at $\nu{=}1/2$. The HR state is an exact zero-energy\footnote{The fact that the HR wave function is a zero mode of the hollow-core Hamiltonian can be seen by noting that when two particles of the same spin separated by a distance $r$ approach each other, the HR wave function vanishes at least as $r^{3}$ (a factor of $r$ from the Fermi-statistics imposed by the determinant and a factor of $r^{2}$ from the $\nu{=}1/2$ bosonic Laughlin intra-spin correlations) while when two particles of opposite spin separated by a distance $r$ approach each other, terms in the expansion of the HR wave function vanish as $r^{0}$ [when the factor of $1/(z^{\uparrow}_{i}-z^{\downarrow}_{j})^{2}$ from the expansion of the determinant is canceled by the corresponding factor in the inter-spin correlations $\prod_{i,j}(z^{\uparrow}_{i}-z^{\downarrow}_{j})^{2}$] or $r^{2}$ (from just the inter-spin correlations) but not as $r^{1}$. Therefore, the HR wave function has no amplitude in the relative angular momentum $m{=}1$ channel and thus has zero energy for the $V_{1}$ Hamiltonian.} state of the hollow-core\footnote{This Hamiltonian is hollow-core since it has $V_{0}{=}0$ and thus does not give any energy penalty to placing two electrons at their closest approach.} Hamiltonian parametrized in terms of Haldane pseudopotentials~\cite{Haldane83} as $\{V_{0}, V_{1}, V_{2}, V_{3}, {\cdots} \}{=}\{0, 1, 0, 0, {\cdots} \}$, where $V_{m}$ is the energy cost of placing two electrons in the relative angular momentum $m$ state. The HR wave function of Eq.~\eqref{eq: Haldane_Rezayi_wf} sans the determinant factor is just the bosonic $\nu{=}1/2$ Laughlin state made up of \emph{all} the particles and the 1/2 Laughlin state is precisely described by the first two terms of the Lagrangian given in Eq.~\eqref{eq:S0}. From here on in, for ease of notation, we shall drop the ubiquitous Gaussian factor from the wave functions.

%%%%%%%%%%%%%%%%%%%%%%%%%%%%%%%%%%%%%%%%%%%%%%%%%%%%%%%%%%%%%%%%%%%%%%%%%%%%%
\begin{figure*}[htpb]
  \begin{subfigure}{0.32\textwidth}
		\includegraphics[width=\textwidth]{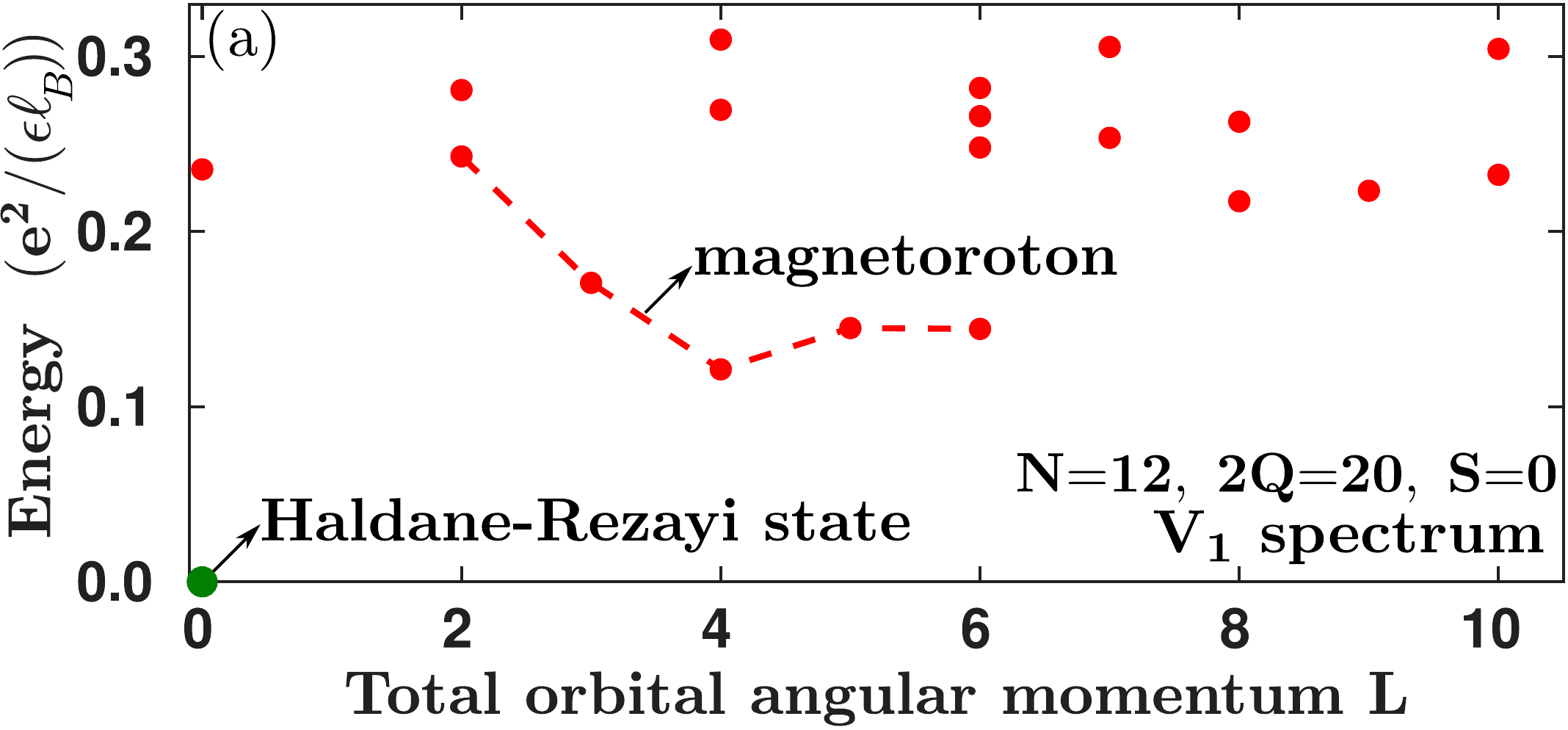}
    \caption{}
  \end{subfigure}
  \begin{subfigure}{0.32\textwidth}
 		\includegraphics[width=\textwidth]{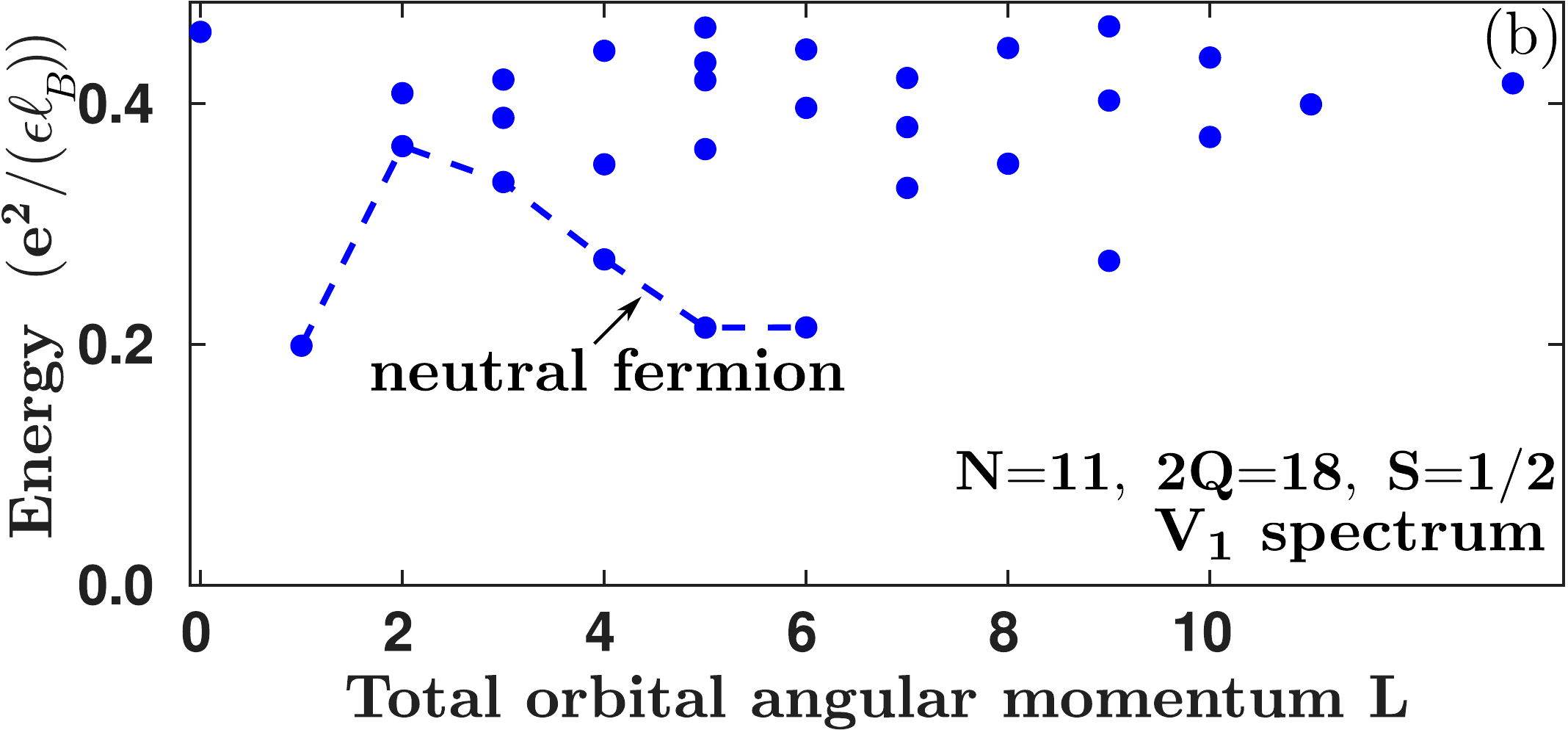}
    \caption{}
  \end{subfigure}
  \begin{subfigure}{0.32\textwidth}
  	\includegraphics[width=\textwidth]{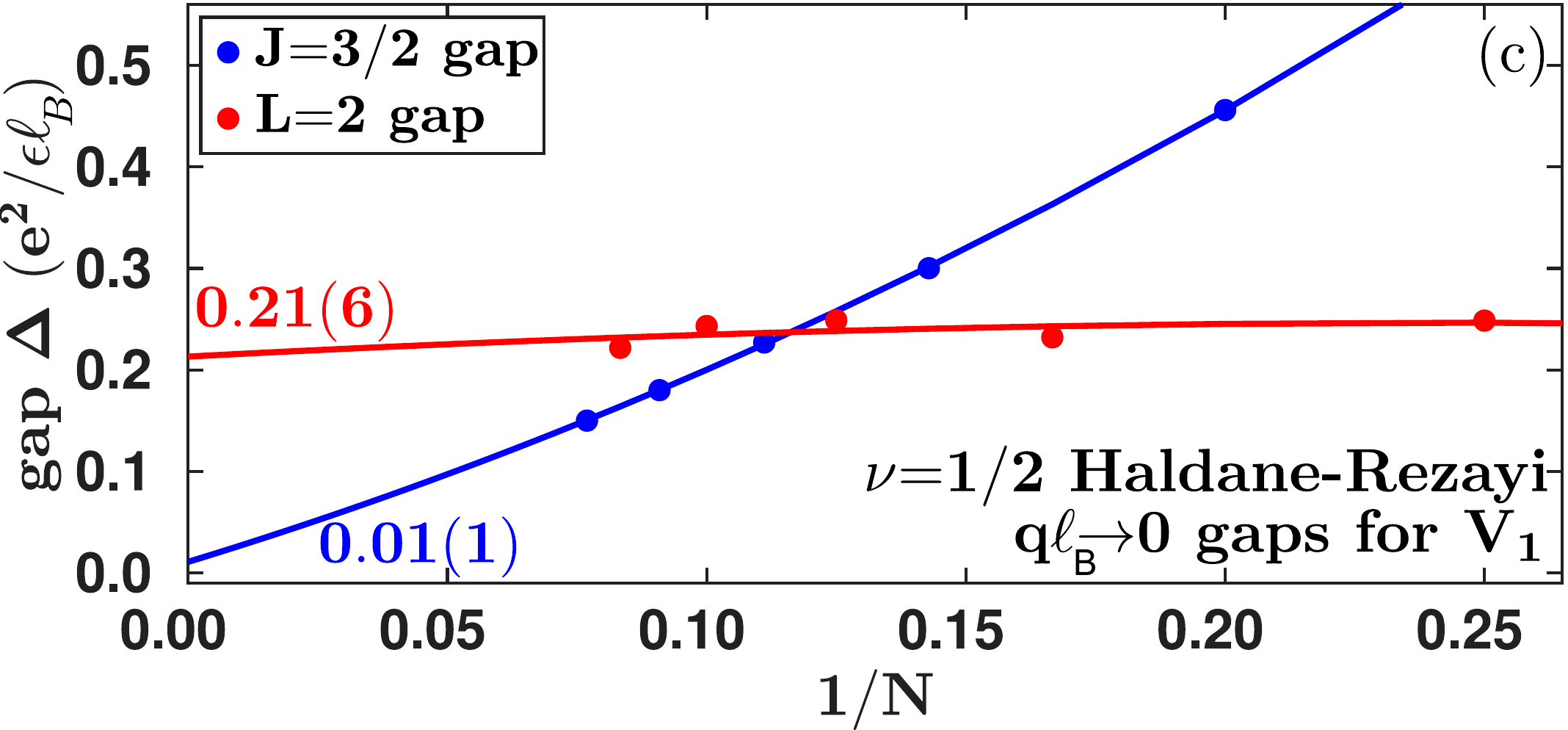}
    \caption{}
  \end{subfigure}
			\caption{(color online) Spectrum and gaps of the collective modes of the $\nu{=}1/2$ Haldane-Rezayi spin-singlet state obtained from exact diagonalization of the $V_{1}$ Haldane pseudopotential Hamiltonian in the spherical geometry for $N$ particles in the presence of $2Q$ flux quanta. The left panel (a) shows the spectrum for an {\it even} number of electrons which supports the Haldane-Rezayi ground state and the real spin $S{=}0$ magnetoroton mode. The center panel (b) shows the spectrum for an {\it odd} number of electrons which supports the real spin $S{=}1/2$ neutral fermion mode. The right panel (c) shows a thermodynamic extrapolation of the long-wavelength limit of the density-corrected magnetoroton and neutral fermion gaps obtained from a quadratic fit in $1/N$. The extrapolated energies are shown on the plot with the number in the parenthesis indicating the error in the intercept.}
		\label{fig: spectra_gaps_Haldane_Rezayi}
\end{figure*}
%%%%%%%%%%%%%%%%%%%%%%%%%%%%%%%%%%%%%%%%%%%%%%%%%%%%%%%%%%%%%%%%%%%%%%%%%%%%%

All our numerical calculations are carried out in the spherical geometry~\cite{Haldane83}. In this geometry, $N$ electrons move on the surface of the sphere at the center of which sits a magnetic monopole that emanates a radial flux of strength $2Qhc/e$. The HR state on the sphere occurs for an even number of electrons $N$ when the flux $2Q{=}2N{-}4$. In Fig.~\ref{fig: spectra_gaps_Haldane_Rezayi}(a) we show the spectrum of the aforementioned hollow-core $V_{1}$-only (strength of which is set to unity as above) Hamiltonian for $N{=}12$ particles.\footnote{Results for smaller systems are similar and the next system of $N{=}14$ which has a Hilbert space dimension of about 4 billion is beyond our reach.} Aside from the zero-energy HR ground state, we can identify a set of low-lying excitations that carry the same spin $S{=}0$ as the HR ground state and form a collective mode analogous to the magnetoroton branch of excitations seen in other FQH states.

Inspired by a recent parton construction of the collective modes of FQH states~\cite{Balram21d}, we propose that the magnetoroton-like mode for the HR state can be described by the wave function
\begin{eqnarray}
  \Psi^{\rm HR-magnetoroton}_{1/2} &=& {\rm Det} \left( \frac{1}{\left(z^{\uparrow}_{i}-z^{\downarrow}_{j}\right)^{2}}\right) \prod_{i,j} \left(z^{\uparrow}_{i} - z^{\downarrow}_{j} \right)^{2} \nonumber \\
	& \times & \Psi^{\rm CFE}_{1/2}( \{z^{\uparrow} \} ) \prod_{i<j} \left(z^{\downarrow}_{i} - z^{\downarrow}_{j}\right)^{2},
	 \label{eq: HR_magnetoroton}
\end{eqnarray}
where $\Psi^{\rm CFE}_{1/2}( \{z^{\uparrow} \} )$ describes the composite fermion~\cite{Jain1} exciton (CFE) mode of the bosonic $\nu{=}1/2$ Laughlin state~\cite{Laughlin:1983fy} made from only the spin-up electrons.\footnote{Since up and down spins are on an equal footing in the HR wave function, one could also create the CFE in spin-down electrons.} This magnetoroton branch of excitations extends from total orbital angular momentum $L{=}2$ to $L{=}N/2$~\cite{Balram16d} [consistent with the identification of the states shown in Fig.~\ref{fig: spectra_gaps_Haldane_Rezayi}~(a)] in the spherical geometry~\cite{Haldane83}. The wavenumber $q\ell_{B}$ on the plane is related to $L$ as $q\ell_{B}{=}L/R$, where $R{=}\sqrt{Q}$ is the radius of the sphere (in units of $\ell_{B}$) and thus the long-wavelength gap of the magnetoroton mode is obtained by looking at its $L{=}2$ gap. The wave function given in Eq.~\eqref{eq: HR_magnetoroton} is amenable to large system evaluations using the Monte Carlo method~\cite{Jain97, Jain97b} in {\it real} space (first quantization). However, it is difficult to evaluate the $V_{1}$ energy of this wave function for large sizes since the $V_{1}$ interaction is not smooth in real space. A different version of the magnetoroton mode can be constructed by replacing the CFE state with the GMP density mode ansatz~\cite{GMP:1985, GMP:1986}. In the long-wavelength limit of our interest, the CFE and GMP versions of the wave function given in Eq.~\eqref{eq: HR_magnetoroton} are identical~\cite{Kamilla96b, Kamilla96c}. The wave function given in Eq.~\eqref{eq: HR_magnetoroton} predicts that the HR graviton has a negative chirality, the same as the chirality of the graviton of the Laughlin states~\cite{Nguyen:2022MultiG, Balram21d,Liou2019}. This is also consistent with the fact that the HR state is annihilated by a short-range Hamiltonian~\cite{Nguyen:2014}.

To look for other low-lying collective modes of the HR state, we calculate the spectrum of the $V_{1}$-only Hamiltonian for an {\it odd} number of electrons at the same ground state flux $2Q{=}2N{-}4$. For an odd number of electrons too, the flux $2Q$ is even and $Q$ is thus integral which produces integral $L$ values for all the states in this Hilbert space. In Fig.~\ref{fig: spectra_gaps_Haldane_Rezayi}(b) we show the $V_{1}$ spectrum for $N{=}11$ electrons.\footnote{Results for smaller systems are similar and the next system of $N{=}13$ which has a Hilbert space of about half a billion is beyond our reach.} In this spectrum, we can identify a set of low-lying states that carry spin $S{=}1/2$ (since the excitation occurs for an odd number of particles, it carries an odd half-integral spin) and form a collective mode which we call the neutral fermion mode in analogy to the neutral fermion branch of excitations seen in the Moore-Read state~\cite{Gunnar:MR, Sreejith11}. The smallest $L$ for the neutral fermion excitation is $L{=}1$ and thus in the long-wavelength limit, it carries an angular momentum of $J{=}3/2$ relative to the ground state (composed of an $L{=}1$ and $S{=}1/2$ excitation and since the mode occurs for $J_{z}{=}L_{z}{+}S_{z}{=}1{+}1/2{=}3/2$ it has $J{=}3/2$ instead of $J{=}1/2$\footnote{For $J{\geq}5/2$, since $S{=}1/2$, $L{\geq}2$ while we clearly see a low-energy state with $L{=}1$ in Fig.~\ref{fig: spectra_gaps_Haldane_Rezayi}(b).}). At the moment, it is not clear how to construct the wave function of the neutral fermion-like mode for the HR state (see \cref{sec: conclusions} for some speculative ideas in that direction). We note here that it might be possible to use the Jack polynomial decomposition of the HR state~\cite{Thomale11} to construct both its magnetoroton and neutral fermion modes~\cite{Yang:MR} in {\it Fock} space (second quantization). However, since this approach is restricted to small systems, with it we will not be able to access the long-wavelength limit of the modes that are of primary focus in the current work.

Finally, we look at the long-wavelength limits of the magnetoroton and neutral fermion gaps which are respectively the $L{=}2$ magnetoroton and $J{=}3/2$ neutral fermion gaps\footnote{Although we have not been able to get the full spectrum for $N{=}13$ we have been able to get the gap of the $J{=}3/2$ excitation for this system and that data point has been included in Fig.~\ref{fig: spectra_gaps_Haldane_Rezayi}(c).}. In Fig.~\ref{fig: spectra_gaps_Haldane_Rezayi}(c) we show a thermodynamic extrapolation of the gaps as a function of $1/N$. Before extrapolation to the thermodynamic limit, the gaps of the finite systems are density corrected~\cite{Morf87}, i.e., the gaps are multiplied by a factor of $\sqrt{2Q\nu/N}$ which corrects for the fact that the density for a finite system on the sphere $N/(4\pi Q\ell_{B}^{2})$ is different from that in the thermodynamic limit $\nu/(2\pi \ell_{B}^{2})$. The density correction weakens the $N$ dependence of the gaps. In the long-wavelength limit, we find that the neutral fermion is gapless (within error bars) which is consistent with previous theoretical expectations~\cite{Read00, Moran:HRGraviton}. On the other hand, the magnetoroton shows a finite gap. A note of caution is warranted here: for spinful systems, there may be stronger finite-size effects than fully polarized systems since only a few systems are amenable to exact diagonalization. For a large system, we expect that in the long wavenumber limit i.e., $q\ell_{B}{\to} \infty$ the neutral fermion and magnetoroton gaps would converge to the same value\footnote{As we suggest in Sec.~\ref{sec: conclusions} the neutral fermion is likely to be composed of a CF-particle (CFP) in one spin and a CF-hole (CFH) in the other spin. The CFE is also made up of a CFP and a CFH but both the excitations reside in the same spin species. In the limit, $q\ell_{B}{\to} \infty$ the CFP and CFH are far away from each other and thus do not interact. Therefore, it does not make a difference as to whether the CFP and CFH carry the same spin (as in the magnetoroton) or have opposite spins (as in the neutral fermion)}. For the systems sizes accessible to us, the long wavenumber limit of the two gaps is still different from each other.

%%%%%%%%%%%%

In the incompressible FQH states such as the Laughlin \cite{Laughlin:1983fy}, Moore-Read \cite{MooreRead,Samkharadze:52}, or Jain states \cite{Jain1}, the entire magnetoroton branch is gapped and its long-wavelength spin-2 excitation acquires a mass term as constructed explicitly in Ref.~\cite{Gromov2017}. However, for the compressible composite fermion Fermi sea state at $\nu{=}1/2$ the magnetoroton is gapless at zero momentum, and thus its graviton is massless \cite{Nguyen:2022MultiG}. In the case of the HR state, although the state is gapless, its magnetoroton could be gapped (as the available numerical results suggest), which implies a massive graviton \cite{Moran:HRGraviton}. One can add the following mass term of the graviton to our action \cite{Gromov2017}
\begin{equation}
\label{eq:mass}
\cS_m=-\int d^3 x\, \sqrt{g}\frac{m}{2}\Bigr(\frac{1}{2}\hat{g}_{ij}g^{ij}-\gamma \Bigl)^2,
\end{equation}
where $g_{ij}$ is the ambient metric, and $\hat{g}_{ij}$ is the emergent metric defined in \cref{eq:metric}. In \cref{eq:mass} $\gamma>2,m>0$ are parameters that determine the graviton mass \cite{Gromov2017,Nguyen2018}. On the other hand, the gravitino seems to remain massless. The mass term of \cref{eq:mass} breaks the supersymmetry as one should not expect that the emergent supersymmetry works exactly. The additional mass term does not change the main conclusions in the previous sections.

\section{Conclusions and outlook}
\label{sec: conclusions}
To summarize, we demonstrated that the $\cN=(1,1)$ supergravity can be realized in the bulk of the Haldane-Rezayi quantum Hall state at filling fraction $\nu=1/2$. We proposed a non-relativistic superalgebra acting on two spatial dimensions. We constructed the CS supergravity action in which the graviton and gravitino are emergent dynamical degrees of freedom that correspond to the low-lying magnetoroton and neutral fermion excitations in the HR state. We were able to reproduce the corresponding edge theory and the long-wavelength limit of the GMP algebra from our proposed model. We also identified the emergent graviton and gravitino excitations by numerical simulation of the HR Hamiltonian on a sphere. Even though our numerical results suggest that the emergent graviton is massive, which suggests the breaking of supersymmetry, for larger systems, the graviton and gravitino may both go soft in the long-wavelength limit. Numerical computations on larger systems sizes, that would give closer access to the long-wavelength limit, are needed to conclusively determine the graviton mass. 

%===========================
We further proposed a trial wave function of the magnetoroton excitation for the HR state in Eq.~\eqref{eq: HR_magnetoroton}. However, the construction of a trial wave function of the neutral fermion mode for the HR state remains an open question. One na\"ive construction would be to use
\begin{eqnarray}
\Psi^{\rm HR-neutral~fermion}_{1/2} &=& {\rm Det} \left( \frac{1}{\left (z^{\uparrow}_{i}-z^{\downarrow}_{j} \right)^{2}}\right) \prod_{i,j} \left(z^{\uparrow}_{i} - z^{\downarrow}_{j} \right)^{2} \nonumber \\
& \times &
\Psi^{\rm CFH}_{1/2}( \{z^{\uparrow} \} ) \Psi^{\rm CFP}_{1/2}( \{z^{\uparrow} \} ).
\label{eq: HR_neutral_fermion}
\end{eqnarray}
Here $\Psi^{\rm CFH}_{1/2}( \{z^{\uparrow} \} )$ describes the composite fermion hole (CFH) of the bosonic $\nu{=}1/2$ Laughlin state made from only the spin-up electrons and $\Psi^{\rm CFP}_{1/2}( \{z^{\downarrow} \} )$ describes the composite fermion particle (CFP) of the bosonic $\nu{=}1/2$ Laughlin state made from only the spin-down electrons (one could also put the CFH in spin-down and CFP in spin-up). The CFH carries an orbital angular momentum of $L^{\rm CFH}{=}(N{+}1)/4$ while the CFP carries $L^{\rm CFH}{=}(N{-}1)/4$\footnote{The CFP and CFH in a Laughlin state made of $N'$ particles carries $L{=}N'/2$~\cite{Balram16d}.}. This neutral fermion branch of excitations would extend from total orbital angular momentum $L{=}3/2$ to $L{=}N/2$ on the sphere and for odd $N$ these orbital angular momenta will be half-integral. However, the flux at which the HR state occurs is $2Q{=}2N{-}4$ is always even (for both even and odd $N$) and thus produces only integral values of $L$. Furthermore, the matrix for which the determinant is evaluated for odd-$N$ is non-square. Thus, this na\"ive construction does not work. This suggests that perhaps one has to modify the determinant factor to construct this excitation by going to a related system with an even number of particles. From exact diagonalization, it appears that the neutral fermion mode extends from $L{=}1$ to $L{=}(N{+}1)/2$. However, it cannot be described by just the magnetoroton of the spin-up electrons (although the maximum $L$ for that matches with $(N{+}1)/2$) since the magnetoroton starts only from $L{=}2$ (the $L{=}1$ exciton is eliminated upon projection to the LLL). We speculate that a construction similar to that carried out by one of us {\it et al.}~\cite{Gromov:SusyMR} for the neutral fermion of the Moore-Read state can be used to construct the wave function of the neutral fermion for the HR state. We can include an additional particle in the spin-down sector to get an effective system with an even number of particles\footnote{Then we have an equal number of up and down spins $N_{\uparrow}{=}N_{\downarrow}{=}(N{+}1)/2$ which would put both spins on an equal footing and also allow for a construction of the determinant of a square matrix as in the HR ground state wave function of Eq.~\eqref{eq: Haldane_Rezayi_wf}.} and then create a CFP in up spin and a CFH in down spin as stated above. By adding the angular momentum of the CFP and CFH, we expect the neutral fermion mode to extend from $L{=}0$ to $L{=}[(N{+}1)/2]/2{+}[(N{+}1)/2]/2{=}(N{+}1)/2$ [this is indeed the largest $L$ up to which we see the mode extend till in Fig.~\ref{fig: spectra_gaps_Haldane_Rezayi}(b)]. The $L{=}0$ state then gets eliminated upon projection to the LLL (like the $L{=}1$ magnetoroton exciton gets projected out). 

The FQH gravitons which are the long-wavelength limit of the magnetoroton excitations can be probed in inelastic light scattering experiments \cite{Pinczuk1, Pinczuk2}. The chirality of the graviton can also be determined with circular polarized Raman scattering \cite{Nguyen:Raman}. In principle, photoluminiscence~\cite{Gunnar:MR} experiments can detect the gravitino excitation, which is the long-wavelength limit of the neutral fermion mode, in the bulk. 

Our results can potentially be extended to shed light on the nature of other unpolarized paired states like the Belkhir-Jain spin-singlet ~\cite{Belkhir93a, Belkhir93} and the Halperin 331~\cite{Halperin83} states that both occur at half-filling and are expected to be fully gapped~\cite{Moran:HRGraviton}. Very recently, by evaluating the energies of the graviton and gravitino excitations, the authors of Ref.~\cite{Pu23} have suggested that the $\nu{=}5/2$ FQH state, modeled by the Moore-Read wave function, is in the vicinity of the proposed supersymmetric point~\cite{Gromov:SusyMR}. A supersymmetric model that captures the bulk physics of the Moore-Read state is still an open question~\cite{Pu23}. Since the proposed supersymmetry of the boundary modes for the Moore-Read state is $\cN=(1,0)$ \cite{Ma:EdgeMR}, one expects that the same supersymmetry should be shared by the bulk theory. Aside from these states, which can all be interpreted as paired states of composite fermions~\cite{Read00}, supersymmetry can be a useful tool to describe the recently proposed $\mathbb{Z}_{n}$-ordered superconducting states of composite bosons~\cite{Balram20}. In general, it appears that paired states of electron-vortex composites can harbor supersymmetry. We expect that the model proposed in this paper can provide some suggestions for further investigations. We also defer the coupling of dynamical emergent graviton and gravitino with the background geometry for future work.

\begin{acknowledgments}
We would like to thank Savdeep Sethi for helpful discussions on two-dimensional supersymmetry. D.X.N. thanks Antal Jevicki, Giandomenico Palumbo, and Patricio Salgado-Rebolledo for fruitful discussions. D.X.N. is supported by grant IBS-R024-D1. K.P. is supported by the NSF grant PHY-2107939 to the University of California, Santa Barbara. A.C.B. thanks the Science and Engineering Research Board (SERB) of the Department of Science and Technology (DST) for funding support via the Start-up Grant No. SRG/2020/000154. Computational portions of this research work were conducted using the Nandadevi supercomputer, which is maintained and supported by the Institute of Mathematical Science's High-Performance Computing Center. Some of the numerical calculations were performed using the DiagHam package, for which we are grateful to its authors. A.G. was supported in part by NSF CAREER Award DMR-2045181, Sloan Foundation, and the Laboratory for Physical Sciences through the Condensed Matter Theory Center. %\note{other grants}
\end{acknowledgments}
%%%%%%%%%%%%%%%%%%%%%%%%%%%%%%%%%%%%%%%%%%%%%%%%%%%

\appendix

%%===========================================================

\section{Dirac algebra and \(\cN = (1,1)\) supersymmetry in two spatial dimensions}
\label{sec:dirac-alg}
In this appendix, we summarize the Dirac algebra and the convention we used in this paper. In the signature \((+,+)\), we consider the Clifford algebra of the gamma matrices
\be
\gamma_a \gamma_b + \gamma_b \gamma_a = -2 \delta_{ab} I, 
\ee
where \(I\) is the \(2 \times 2\) identity matrix and $\delta_{ab}$ is the metric in a Cartesian basis. We also define $\hat{\gamma}$ as
\be
\hat\gamma = \tfrac{1}{2} \varepsilon^{ab} \gamma_a \gamma_b
\ee
where $\varepsilon_{ab}$ is the totally antisymmetric tensor in two dimensions. It will be convenient to use the following Weyl (chiral) form of the Dirac algebra which diagonalizes \(\hat\gamma\):
\be
\label{eq:gamma}
\gamma_x = i \sigma_y = \begin{pmatrix}
		0 & 1 \\ -1 & 0
	\end{pmatrix} &\eqsp \gamma_y = i \sigma_x = \begin{pmatrix}
		0 & i \\ i & 0
	\end{pmatrix} \\ \hat\gamma = i \sigma_z &= \begin{pmatrix}
		i &0 \\ 0 & -i
	\end{pmatrix}
\ee
where \(\sigma_i\) are the standard Pauli matrices.

A spinor \(\psi^\alpha\) is given by a two-component matrix
\be
\psi^\alpha = \begin{pmatrix}\psi^1 \\ \psi^2 \end{pmatrix} ,
\ee
If the Dirac matrices are chosen as in \cref{eq:gamma} then the Majorana condition on the spinor is \(\psi^2 = \overline{\psi^1}\). Spinor indices can be raised and lowered with the antisymmetric symbols \(\epsilon_{\alpha \beta}\) and \(\epsilon^{\alpha\beta}\).\\

Next, we show that the algebra in \cref{eq:algebra0} is equivalent to the \(\cN = (1,1)\) supersymmetry algebra. Take the holomorphic complex coordinate on the plane as \(z = x + iy\) and its complex conjugate. The holomorphic/antiholomorphic translations are then
\be
  \cP_z = \frac{1}{2}(\cP_x -i \cP_y) \eqsp \cP_{\bar z} = \overline{\cP_z}.
\ee
Similarly, decompose the rotation into chiral holomorphic/antiholomorphic components \(\mc J = i (J - \bar J)\). The supersymmetry generator is a Majorana spinor operator \(Q_\alpha = (Q, \bar Q)\). In terms of these, using \cref{eq:gamma}, \cref{eq:algebra0} is equivalent to the chiral holomorphic algebra
\be
\label{eq:algebra}
\lb[J,\cP_z\rb] = \cP_z \eqsp \lb[Q,Q\rb] = 2 \cP_z \eqsp \lb[J,Q\rb] = \tfrac{1}{2} Q,
\ee
and its complex conjugate. Now identifying \(J = L_0\), \(\cP_z = L_{-1}\) and \(Q = G_{-1/2}\), we see that \cref{eq:algebra} is the subalgebra of global holomorphic supersymmetries within the Neveu-Schwarz superalgebra (see \cite{GSW, FMS}). Including the complex conjugate algebra then gives us the full global \(\cN = (1,1)\) supersymmetries in two dimensions.\footnote{Note that in a two-dimensional CFT the bosonic symmetry generators can be extended to the Virasoro algebra and its supersymmetric extension is the Neveu-Schwarz algebra.}
%%%%%%%%%%%%%%%%%%%%%%%%%%%%%%%%%%%%%%%%
\section{Derivation of the action}
\label{sec:derivation}
In this section, we will explicitly derive the action \eqref{eq:Ls1} using the algebra \eqref{eq:algebra0} and the bilinear invariants \eqref{eq:invariants}. We use the complex vielbein basis 
    \begin{equation}
	\hat{e}_\mu=\frac{1}{2}\lb(\he^x_\mu+i \he^y_\mu \rb), \quad \bar{\he}_\mu=\frac{1}{2}\lb(\he^x_\mu-i \he^y_\mu \rb)
\end{equation}
to rewrite the connection $\cB$
\begin{equation}
	\cB_\mu=\hw_\mu \cJ +\he_\mu \cP_z+\bar{\he}_\mu \cP_{\bar z}  +\bar{\psi}_\mu \bar{Q}+\psi_\mu Q
\end{equation}
We work expand  explicitly the last term of the CS action \eqref{eq:Ls0}
\begin{equation}
\label{eq:BBB}
	\text{sTr}\lb(\cB \wedge \cB \wedge \cB\rb)=\text{sTr}\lb(\epsilon^{\mu\nu\lambda} \cB_\mu \cB_\nu \cB_\lambda\rb)
\end{equation}
Let's consider the terms with $\cJ, P, \cP_{\bar z}$ in \eqref{eq:BBB}
\be
	\label{eq:JPPb}
	\text{sTr}[\epsilon^{\mu\nu\lambda} \hw_\mu \cJ\left(\he_\nu \cP_z \bar{\he}_\lambda \cP_{\bar z}+\bar{\he}_\nu \cP_{\bar z} \he_\lambda \cP_z \right) \\ +\epsilon^{\mu\nu\lambda} \bar{\he}_\mu \cP_{\bar z}\left(\he_\nu \cP_z \hw_\lambda \cJ+\hw_\nu \cJ \he_\lambda \cP_z \right)  \\ +\epsilon^{\mu\nu\lambda} \he_\mu \cP_z\left(\bar{\he}_\nu \cP_{\bar z} \hw_\lambda \cJ+\hw_\nu \cJ \bar{\he}_\lambda \cP_{\bar z} \right) ]
\ee
Using the property of the super trace \eqref{eq:str}, the first term of \eqref{eq:JPPb} can be written as 	
\be
	\text{sTr}& \lb[ \epsilon^{\mu\nu\lambda} \left(\hw_\mu \cJ\he_\nu \cP_z \bar{\he}_\lambda \cP_{\bar z}+\he_\lambda \cP_z\hw_\mu \cJ\bar{\he}_\nu \cP_{\bar z}  \right) \rb]\\&=\text{sTr} \lb[ \epsilon^{\mu\nu\lambda} \left(\hw_\mu \cJ\he_\nu \cP_z \bar{\he}_\lambda \cP_{\bar z}-\he_\nu \cP_z\hw_\mu \cJ\bar{\he}_\lambda \cP_{\bar z}  
	\right) \rb] \\
	&=\text{sTr} \lb[ \epsilon^{\mu\nu\lambda} \hw_\mu \he_\nu \bar{\he}_\lambda\left[ \cJ ,\cP_z  
	\right] \cP_{\bar z}\rb]\\&=\epsilon^{\mu\nu\lambda} \hw_\mu \he_\nu \bar{\he}_\lambda\text \, {sTr} \lb[ i \cP_z \cP_{\bar z}\rb] \\&=i \mu_2 \epsilon^{\mu\nu\lambda} \hw_\mu \he_\nu \bar{\he}_\lambda= i \mu_2 \hw \wedge \he \wedge \bar{\he},
\ee
where we used the commutation relation of $\cJ$ and $P$ and the bilinear invariants \eqref{eq:invariants}. Similarly, the second term and the last term of \eqref{eq:JPPb} give the same results 
\begin{align}
  \!\!\!\!\!\! \text{sTr} \lb[ \epsilon^{\mu\nu\lambda} \bar{\he}_\mu \cP_{\bar z}\left(\he_\nu \cP_z \hw_\lambda \cJ+\hw_\nu \cJ \he_\lambda \cP_z \right)\rb]=i \mu_2 \hw \wedge \he \wedge \bar{\he} \\
\!\!\!\!\!\!	\text{sTr} \lb[ \epsilon^{\mu\nu\lambda} \he_\mu \cP_z\left(\bar{\he}_\nu \cP_{\bar z} \hw_\lambda \cJ+\hw_\nu \cJ \bar{\he}_\lambda \cP_{\bar z} \right)\rb]=i \mu_2 \hw \wedge \he \wedge \bar{\he}
\end{align}
Consequently, the terms with $\cJ, \cP_{z}, \cP_{\bar z}$ of \eqref{eq:BBB} gives us 
\begin{equation}
	i 3 \mu_2 \hw \wedge \he \wedge \bar{\he}.
\end{equation}
Subsequently, we consider the terms with $\cP_z, \bar{Q}, \bar{Q}$ in \eqref{eq:BBB}

\be
	\label{eq:PQbQb}
	\text{sTr} [ \epsilon^{\mu\nu\lambda} \he_\mu \cP_{z}\bar \psi_\nu \bar Q \bar{\psi}_\lambda \bar{Q}+\epsilon^{\mu\nu\lambda} \bar{\psi}_\mu \bar{Q}\bar \psi_\nu \bar Q \he_\lambda \cP_{z} \\+\epsilon^{\mu\nu\lambda} \bar{\psi}_\mu \bar{Q}\he_\nu \cP_{z} \bar \psi_\lambda \bar Q  ]
\ee
We use the anti-commutation of Grassmannian fields to rewrite the first term of \eqref{eq:PQbQb} as 
\be
	\text{sTr} \lb[ \epsilon^{\mu\nu\lambda} \he_\mu \cP_{z}\bar \psi_\nu \bar Q \bar{\psi}_\lambda \bar{Q}\rb]=\epsilon^{\mu\nu\lambda}\he_\mu \bar \psi_\nu \bar{\psi}_\lambda  \text{sTr} \lb[\cP_{z} \frac{1}{2}[\bar Q, \bar Q] \rb]\\
 =\epsilon^{\mu\nu\lambda}\he_\mu \bar \psi_\nu \bar{\psi}_\lambda  \text{sTr} \lb[\cP_{z}  \cP_{\bar z}\rb]
 =\mu_2\epsilon^{\mu\nu\lambda}\he_\mu \bar \psi_\nu \bar{\psi}_\lambda =\mu_2 \he \wedge \bar \psi \wedge \bar \psi
\ee
The second term and the last term of \eqref{eq:PQbQb}
give similar results. Consequently, the terms with $\cP_{z}, \bar{Q}, \bar{Q}$ in \eqref{eq:BBB} gives
\begin{equation}
	3 \mu_2 \he \wedge \bar \psi \wedge \bar \psi.
\end{equation}
Similarly, the terms with $\bar \cP_{z}, Q ,Q$ in \eqref{eq:BBB} gives 
\begin{equation}
	3 \mu_2 \bar \he \wedge \psi \wedge  \psi.
\end{equation}
Finally, we consider the terms with $\cJ, Q, \bar{Q}$ in \eqref{eq:BBB}
\be
	\label{eq:JQQb}
	\text{sTr}&[ \epsilon^{\mu\nu\lambda} \hw_\mu \cJ\left(\psi_\nu Q \bar{\psi}_\lambda \bar{Q}+\bar{\psi}_\nu \bar{Q} \psi_\lambda Q \right)\\&+\epsilon^{\mu\nu\lambda} \bar{\psi}_\mu \bar{Q}\left(\psi_\nu Q \hw_\lambda \cJ+\hw_\nu \cJ \psi_\lambda Q \right)\\ &+\epsilon^{\mu\nu\lambda} \psi_\mu Q\left(\bar{\psi}_\nu \bar{Q} \hw_\lambda \cJ+\hw_\nu \cJ \bar{\psi}_\lambda \bar{Q} \right)]
\ee
The first term of \eqref{eq:JQQb} can be manipulated as 
\be
	\text{sTr}& \lb[ \epsilon^{\mu\nu\lambda}\left( \hw_\mu \cJ\psi_\nu Q \bar{\psi}_\lambda \bar{Q}+\hw_\mu \cJ\bar{\psi}_\nu \bar{Q} \psi_\lambda Q \right) \rb] \\&=	\text{sTr} \lb[ \epsilon^{\mu\nu\lambda}\left( \hw_\mu \cJ\psi_\nu Q \bar{\psi}_\lambda \bar{Q}+\psi_\lambda Q\hw_\mu \cJ\bar{\psi}_\nu \bar{Q}  \right) \rb]\\
	&=\epsilon^{\mu\nu\lambda} \hw_\mu \psi_\nu \bar \psi_\lambda	\text{sTr} \lb[  [\cJ,Q] \bar{Q}  \rb]\\&=\epsilon^{\mu\nu\lambda} \hw_\mu \psi_\nu \bar \psi_\lambda\text{sTr} \lb[ \frac{i}{2}Q \bar{Q}  \rb]=-\frac{\mu_3}{2} \hw \wedge \psi \wedge \bar \psi
\ee
where we used \eqref{eq:str} and the anti-commutation of Grassmannian fields. The other terms of \eqref{eq:JQQb} give the same results. Consequently, the terms with $\cJ, Q, \bar{Q}$ in \eqref{eq:BBB} give
\begin{equation}
	-\frac{3\mu_3}{2} \hw \wedge \psi \wedge \bar \psi. 
\end{equation}
Using the invariants \eqref{eq:invariants} and equation \eqref{eq:Bmu}, we can easily obtain 
\be
 \text{sTr}\lb(\cB \wedge d \cB\rb)=\mu_1 \hw d \hw+\frac{\mu_2}{2} \delta_{ab}\he^a d \he^b+ i\mu_3 \epsilon_{\alpha \beta}\psi^\alpha d \psi^\beta
\ee
After converting the terms in  $\text{sTr}\lb(\cB \wedge \cB \wedge \cB\rb)$ to the normal coordinate basis, we have the following result
\be
    \text{sTr}\lb(\cB \wedge d \cB+\frac{3}{2}\cB \wedge \cB \wedge \cB\rb)= \mu_1 \hw d \hw+ \frac{\mu_2}{2} \delta_{ab}\he^a d \he^b\\
	+\mu_2  \he^a\wedge\lb(\frac{\hw}{2} \epsilon_{b a}\wedge \he^b+\delta_{ab}\gamma^b_{\alpha\beta}\psi^\alpha \wedge \psi^\beta    \rb)\\+ i\mu_3 \epsilon_{\alpha \beta}\psi^\alpha d \psi^\beta+i \frac{\mu_3}{2}  \psi^\alpha \epsilon_{\alpha\beta}\wedge \hw \hat{\gamma}^\beta{}_\gamma \wedge \psi^\gamma 
\ee
Using the definition of the torsion \eqref{eq:torsion}, we obtain explicitly the terms with coefficients $\mu_1, \mu_2$, and $\mu_3$ in the action \eqref{eq:Ls1} in the main text.

%%%%% END
\newpage
\bibliography{SG}
\end{document}